\documentclass[runningheads]{llncs}
\usepackage{eccv}

\usepackage{eccvabbrv}
\usepackage{graphicx}
\usepackage{booktabs}
\usepackage{textgreek}
\usepackage{tikz}
\usepackage{amssymb}
\usepackage{multirow}
\usepackage{wrapfig}
\usepackage{soul}
\usepackage{tikz}
\usepackage[dvipsnames]{xcolor}

\usepackage[accsupp]{axessibility}  

\usepackage[pagebackref,breaklinks,colorlinks,citecolor=eccvblue]{hyperref}
\definecolor{bole}{rgb}{0.47, 0.27, 0.23}
\definecolor{goldenpoppy}{rgb}{0.99, 0.76, 0.0}
\definecolor{goldenyellow}{rgb}{1.0, 0.87, 0.0}
\definecolor{palesilver}{rgb}{0.79, 0.75, 0.73}
\definecolor{bronze}{rgb}{0.8, 0.5, 0.2}

\definecolor{fuzzywuzzy}{rgb}{0.8, 0.4, 0.4}
\definecolor{indigo}{rgb}{0.29, 0.0, 0.51}
\definecolor{dark_green}{rgb}{0, 0.0, 0}
\newcommand{\ky}[1]{{\color{dark_green}#1}}

\newcommand{\gs}[1]{{\color{black}#1}}

\renewcommand{\paragraph}[1]{\noindent\textbf{$\blacktriangleright$ #1}}

\newcommand{\q}{\mathcal{Q}}
\newcommand{\f}{\mathcal{F}}
\newcommand{\ray}{\mathbf{r}}
\newcommand{\x}{\mathbf{x}}
\newcommand{\dir}{\mathbf{d}}
\newcommand{\dfield}{\mathcal{D}}
\newcommand{\parsNerf}{\boldsymbol{\theta}}
\newcommand{\parsDef}{\boldsymbol{\theta}_\dfield}
\newcommand{\parsQuad}{\boldsymbol{\theta}_\f}
\newcommand{\expect}{\mathbb{E}}
\newcommand{\mesh}{\mathcal{M}}
\newcommand{\vertex}{\mathcal{V}}
\renewcommand{\triangle}{\mathcal{T}}
\newcommand{\iView}{c}
\newcommand{\gc}{\tikz\draw[black,fill=goldenyellow] (0,0) circle (.6ex);}
\newcommand{\silc}{\tikz\draw[black,fill=palesilver] (0,0) circle (.6ex);}
\newcommand{\brzc}{\tikz\draw[black,fill=bronze] (0,0) circle (.6ex);}

\usepackage{orcidlink}

\begin{document}

\title{Volumetric Rendering\\ with Baked Quadrature Fields} 

\titlerunning{Quadrature Fields}
\author{
  Gopal Sharma\orcidlink{0000-0002-7492-7808}\textsuperscript{1},
  Daniel Rebain\textsuperscript{1},
  Kwang Moo Yi\orcidlink{0000-0001-9036-3822}\textsuperscript{1},
  Andrea Tagliasacchi\orcidlink{0000-0002-2209-7187}\textsuperscript{2, 3, 4}
}
\institute{
  \textsuperscript{1} University of British Columbia, 
  \textsuperscript{2} Google DeepMind,\\
  \textsuperscript{3} Simon Fraser University,
  \textsuperscript{4} University of Toronto
  \url{https://quadraturefields.github.io}
  }
\authorrunning{G.~Sharma, D.~Rebain, K.M.~Yi and A.~Tagliasacchi}



\maketitle

\begin{figure}[]
    \centering
\includegraphics[width=\linewidth]{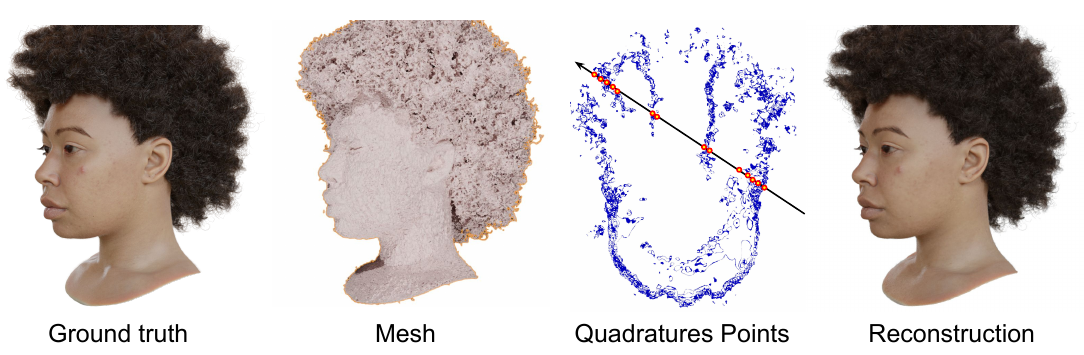}
\caption{We propose using textured polygons with NeRF to efficiently render non-opaque scenes, combining high-quality rendering with modern graphics hardware. 
To model a scene, we produce a mesh that gives quadrature points along a ray~(shown as points on the intersection with the cross-section of the mesh) required in volumetric rendering.
}\label{fig:teaser}
\end{figure}
\begin{abstract}
We propose a novel Neural Radiance Field (NeRF) representation for non-opaque scenes that enables fast inference by utilizing textured polygons.
Despite the high-quality novel view rendering that NeRF provides, a critical limitation is that it relies on volume rendering that can be computationally expensive and does not utilize the advancements in modern graphics hardware.
Many
existing methods fall short when it comes to modelling volumetric effects as they rely purely on surface rendering.
We thus propose to model the scene with polygons, which can then be used to obtain the quadrature points required to model volumetric effects, and also their opacity and colour from the texture.
To obtain such polygonal mesh, we train a specialized field whose zero-crossings would correspond to the quadrature points when volume rendering, and perform marching cubes on this field.
We then perform ray-tracing and utilize the ray-tracing shader to obtain the final colour image.
Our method allows an
easy integration with existing graphics frameworks allowing rendering speed of over 100 frames-per-second for a $1920\times1080$ image, while still being able to represent non-opaque objects.
\end{abstract}

\section{Introduction}
Neural Radiance Fields (NeRFs) \cite{Mildenhall2020nerf} have gained popularity by demonstrating impressive capabilities in generating photo-realistic novel views. 
They use a continuous volumetric function to represent a scene 
with
a 5D implicit function that estimates
the density and radiance for any position and direction. 
NeRF representations are trained to achieve multi-view colour consistency for a set of posed images.
\gs{One of the main challenges to the widespread adoption of NeRF is the high computation cost.}
For example, the traditional implementation of NeRF involves a volumetric rendering algorithm that calculates the density and radiance by evaluating a large MLP at hundreds of sample positions along the ray for each pixel. 
This rendering process is too slow for interactive visualization without powerful GPUs.

Researchers have thus been exploring real-time rendering methods using voxel grids \cite{snerg2021} and polygonal meshes \cite{chen2022mobilenerf,yariv2023bakedsdf} to address the challenge of representing scene geometry in neural volumetric rendering. 
While MobileNeRF~\cite{chen2022mobilenerf} and BakedSDF~\cite{yariv2023bakedsdf} have made progress in using \textit{binary} opacities to restrict volumetric content to polygonal meshes they cannot represent \textit{transparent} surfaces such as glass or clouds as they rely on a \textit{single} point to render each ray, \gs{thus} unable to represent anything other than hard surfaces.
To overcome this limitation, multiple quadrature points need to be sampled along a ray within the neural volumetric rendering setup.\footnote{Note that simply using multiple quadrature points does not enable modeling complicated physics-based rendering such as refraction, but we limit ourselves in this work to what is possible strictly with volume rendering.}
\gs{ 
Towards that end, MERF~\cite{reiser2023merf} allows for fast rendering of large-scale scenes with a small memory footprint by utilizing a sparse feature grid and high-resolution 2D feature planes.
They use a hybrid of planar and volumetric representations to model the scene but do not extract an explicit geometry that can enable downstream tasks in graphics and simulation.}
Thus, precisely and efficiently storing quadrature points needed for volumetric rendering is difficult and still an open problem.
Partially circumventing this problem, Adaptive-Shells~\cite{adaptiveshells2023} enables the rendering of fur and hair by performing volumetric rendering within a shell defined around the surface extracted from a signed distance field.
However, their inherent representation based on SDF can be limiting in glassy media where SDF fails to recover a correct surface.

\gs{\ky{In this work, our goal} is to ``bake'' an existing NeRF model capable of approximating both solid and transparent objects, while still being capable of running in real-time.
Toward this objective, we start from the key insight that mesh intersection-finding algorithms are highly efficient in modern graphics hardware.
\ky{Thus, if we were to have a mesh that, for each ray, the intersection points of the ray with this mesh correspond to the points that are supposed to be used for volume rendering, then the rendering process can be made highly efficient.}
More specifically, we aim for a mesh that on hard surfaces creates a single intersection point with the ray thus imitating surface rendering, and at transparent surfaces that require volume rendering creates multiple intersection points, \eg, through wrinkles.
While the insight is straightforward, implementing a method that can find such a surface is non-trivial.
}

To implement this insight we propose learning an auxiliary neural field whose zero crossing surfaces induce a set of quadrature points for NeRF volumetric rendering -- we name this field the \emph{quadrature field}.
We train the quadrature field so that it aligns with the \textit{surface-field}~\cite{goli2022nerf2nerf} of the scene being represented; see~\cref{fig:onion}.
\footnote{Surface fields apply to both solid and transparent objects.}
With this field, we use marching cubes to extract the polygonal mesh, and for each ray, we use the intersection points with the mesh as quadrature points for volume rendering.
To train this field, we use its gradients to encourage quadrature points to occur near 
the surfaces, as shown in the~\cref{fig:teaser}.

We evaluate our method on several datasets, including the Shelly dataset consisting of non-opaque shapes~\cite{adaptiveshells2023} and the MipNeRF-360 dataset of real scenes.
To summarize, our contributions are as follows:
\begin{itemize}
\item \gs{our approach produces volumetric effects and successfully reconstructs non-solid objects while using only triangular mesh to place quadrature points;}
\item \ky{we show that by doing so, one can take full leverage of modern graphics hardware, achieving over 100 frames-per-second for rendering $1920\times1080$ images;}
\item we introduce a quadrature field and use it to train a neural field to extract quadrature points for both solid and transparent objects;
\item our approach produces a compact representation of the scene and produces comparable results to the originally trained NeRF models.
\end{itemize}
\section{Related works}
Novel view synthesis has been studied extensively in the literature~\cite{SOTArendering}. In this section, we review previous works with a focus on real-time rendering.

\paragraph{Light fields}
When viewpoints are densely sampled, novel-view synthesis can be achieved through light field rendering~\cite{Levoy96}.
Lumigraph~\cite{lumigraph} performs interpolation between observed rays for rendering novel views, but this approach demands significant memory and restricts camera movements.
These challenges can be mitigated by utilizing optical flow~\cite{OmniPhotos} for image interpolation or by employing neural networks to represent light fields~\cite{attal2022learning}. 
Multi-plane~\cite{Soft3DReconstruction, Flynn2019, mildenhall2019llff} and multi-sphere image representations~\cite{attal6d0f} have demonstrated usefulness, although they still limit camera movement.
However, in practical settings where observed viewpoints are not densely captured, reconstructing a 3D representation of the scene is crucial for rendering convincing novel views.

\paragraph{Mesh rendering (classical)}
Traditional approaches to generating novel views utilize triangle meshes, typically reconstructed from point clouds via a multi-step process involving multi-view stereo~\cite{Furukawa2015}, Poisson surface reconstruction~\cite{psr2006}, and marching cubes~\cite{marchingcubes}.
To create novel views, observed images are re-projected into each desired viewpoint and merged using either predetermined~\cite{lumigraph2023} or learned blending weights~\cite{freeviewhedman2018,Riegler2020FVS,Riegler2021SVS}.
Although mesh-based representations are suitable for real-time rendering, they often exhibit inaccurate geometry in regions with intricate details or complex materials, resulting in visible imperfections.

\paragraph{Mesh rendering (differentiable)}
It is also possible to compute explicit triangle meshes through differentiable inverse rendering.
For instance, DefTet~\cite{gao2020deftet} differentiably renders a tetrahedral grid, considering occupancy and colour at each vertex, and then composes the interpolated values along a ray.
NVDiffRec~\cite{Munkberg_2022_CVPR} combines differentiable marching tetrahedra~\cite{shen2021dmtet} with differentiable rasterization to perform full inverse rendering, allowing extraction of triangle meshes, materials, and lighting from images. 
While these approaches enable scene editing and relighting, they tend to compromise view synthesis quality.

\paragraph{Neural radiance fields~(NeRF)}
Neural radiance fields \cite{Mildenhall2020nerf} learn 3D consistent scene representation using continuous opacities and radiance with the help of an MLP.
This approach has produced excellent results for the novel-view synthesis of 3D objects with reflections \cite{verbin2022refnerf}, outdoor bounded scenes \cite{barron2022mipnerf360} and unbounded scenes \cite{rematas2022urf}.
However, as volumetric rendering produces pixel colour by evaluating the MLP over hundreds of points per ray, rendering speed is limited.

\paragraph{Efficient rendering of NeRFs}
There have been several ways to speed up training and inference of NeRF.
Early attempts include AutoInt~\cite{weiping2018autoint}, which models the radiance and opacity of a segment of ray instead of individual points and alpha composite the values over segments to get pixel colour, and DeRF~\cite{derf}, which combines multiple small NeRFs trained to represent disjoint spaces.
Garbin \etal~\cite{Garbin2021FastNeRFHN} introduced caching a factorized representation of neural radiance field for fast inference albeit at higher memory requirement for the cache.
More recently, this problem has most commonly been addressed by trading off compute vs. storage by storing features into grids.
These feature grids can be dense voxel grids~\cite{yu_and_fridovichkeil2021plenoxels}, sparse voxel grids~\cite{snerg2021}, multi-resolution hash grids~\cite{mueller2022instant}, small MLPs distributed spatially~\cite{kilonerf} and low-rank tensor approximations of dense grids~\cite{Chen2022ECCV}.
While in these methods features are converted into radiance/density by a small MLP, diffuse colours can also be stored on the grid and view-dependent radiance be represented by spherical harmonics~\cite{yu2021plenoctrees, yu_and_fridovichkeil2021plenoxels, ReluField_sigg_22}.
Recently, Kerbl~\etal \cite{kerbl3Dgaussians} proposed learning a sparse set of 3D Gaussians to represent the scene, which allows skipping the empty regions of the scene easily and gives real-time rendering.
\gs{While their approach can also reconstruct volumetric media, their primitives are disconnected, hindering several applications in graphics and simulations.}

\paragraph{Baking neural features}
Rather than accelerating the NeRF directly, one can also ``bake'' the neural features into polygonal meshes or volumetric textures.
SNERG \cite{snerg2021} proposed to store features in sparse volumetric textures, and volumetric ray marching combined with deferred rendering to generate pixel colours. 
However, this approach 
requires a large amount of GPU memory to store volumetric data.
Recently, MERF~\cite{reiser2023merf} allows for fast rendering of large-scale scenes while utilizing smaller memory in comparison to SNERG by utilizing a sparse feature grid and high-resolution 2D feature planes.
They use a hybrid of planar and volumetric representations to model the scene, which is orthogonal to our approach which is purely surface-based.
In contrast, MobileNeRF~\cite{chen2022mobilenerf} proposed using a classical triangular mesh for baking the neural features into a texture map, and used \textit{binary} opacities to optimize for a rasterized representation via volumetric rendering.
BakedSDF~\cite{yariv2023bakedsdf} starts with VolSDF to learn a surface-based neural radiance field and use learned features baked at mesh vertices for real-time rendering.
This produces smoother and cleaner meshes in comparison to MobileNeRF owing to SDF priors via Eikonal loss and leads to better performance.
However, both MobileNeRF and BakedSDF are unable to represent transparent objects faithfully as both representations assume rays to terminate at the first intersection with a surface.

\paragraph{Modeling transparent scenes}
$\alpha$Surf~\cite{wu2023alphasurf} reconstructs 3D geometry of semi-transparent objects such as glass.
Their approach is based on a field that is initialized using normalized values of volumetric density, in comparison to our approach which is based on the surface field derived from volumetric weights.
Their approach extracts faithful surfaces for transparent scenes but they do not explore the accurate placement of these surfaces for novel-view synthesis applications. 
NEMTO~\cite{wang2023nemto} models transparent objects by extracting the geometry of the object using a DeepSDF~\cite{Park_2019_CVPR} based approach and using another network to model bending of the ray through transparent media.
Their reliance on the SDF-based representation prevents them from modeling volumetric effects like hair and fur, which our models can model easily as shown in~\cref{sec:experiments}.
\gs{Most recently, Adaptive-shells~\cite{adaptiveshells2023} performs sample efficient volumetric rendering by modifying the NeUS~\cite{wang2021neus} representation by predicting the scale parameter to adapt to the local volumetric media.
The scale is used to extract a shell of triangular mesh which constrains the region where volumetric rendering is done with the help of an Instant-NGP~\cite{mueller2022instant} backbone.
Their reliance on an SDF as the intrinsic representation can be limiting in approximating a glass-like media.
In contrast, we extract the mesh using a general quadrature field (see~\cref{sec:q-field}) and define volumetric rendering by finding the quadrature points via ray-mesh intersection. This allows our approach to represent glass-like objects as shown in the~\cref{fig:bottled-ship}.
}

\section{Method}
Given a set of posed images, our goal is to create a compact 3D representation of the scene that allows fast rendering.
Similarly to MobileNeRF~\cite{chen2022mobilenerf}, the representation consists of a triangular mesh and a texture map consisting of neural features and continuous opacities.
Our rendering process consists of two steps:
\begin{enumerate}
    \item \gs{We use ray tracing to render a transparent scene (i.e. continuous opacities) by alpha compositing colour at points of intersection of the ray and the triangular mesh; see \ref{sec:baking}.}
    \item We render view-dependent effects via spherical Gaussian lobes stored in the texture-map;
\end{enumerate}
\begin{figure*}[t]
\centering
\includegraphics[width=\linewidth, trim=0 0 0 0]{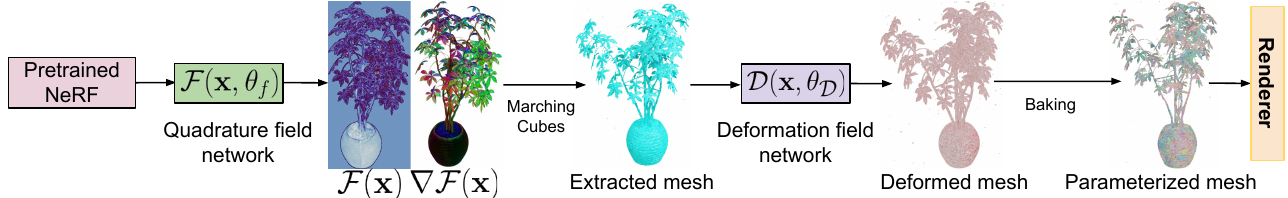}
\caption{\textbf{Overview of our pipeline.}
We start with a pre-trained network to train a quadrature field that learns the placement of quadrature points.
The extracted mesh from the quadrature field is fine-tuned using a deformation field (deformation is shown using red colour on the deformed mesh). 
Lastly, the neural features are baked into a texture map and the mesh, which can be rendered \gs{using ray-tracing}.}
\label{fig:pipeline}
\end{figure*}
\noindent
This representation is created in four-stages (also illustrated in the~\cref{fig:pipeline}):
\begin{enumerate}
    \item \textit{Training the NeRF.} We train a NeRF model with continuous opacities in which quadrature points are sampled using importance sampling~(\cref{sec:nerf}).
    \item \textit{Training the quadrature-field.} We train the quadrature field network with the help of NeRF. We use the trained quadrature field to extract a mesh~(\cref{sec:q-field}).
    \item \textit{Fine-tuning.} We 
    fine-tune the mesh vertices and 
    NeRF 
    with a network that produces the deformation field~(\ref{sec:finetuning}).
    \item \textit{Baking.} 
    We extract the neural features on the surface of the mesh and bake these features into a texture-map~(\cref{sec:baking}).
\end{enumerate}

\subsection{Training the NeRF}
\label{sec:nerf}
Mildenhall~\etal~\cite{Mildenhall2020nerf} introduced a 3D scene representation consisting of an MLP with trainable parameters $\parsNerf$ that takes a position $\mathbf{x} \in \mathbb{R}^3$ and a direction vector $\mathbf{d}$ and outputs radiance $\mathbf{c}(\mathbf{x}, \mathbf{d})$ and density~$\sigma(\mathbf{x})$.
Given a camera pose, the pixel colour is computed using volumetric integration along a ray $\mathbf{r} = (\mathbf{o}, \dir)$ which is sampled at quadratures~$\{t_i\}$ inducing a set of spatial samples $\x_i = \mathbf{o} + t_i\mathbf{d}$:
\begin{equation}
    \mathbf{C}(\ray; \parsNerf) = \sum_i w_i \cdot c_i
    \label{eq:rendering}
\end{equation}
where $w_i = \alpha_i\Pi_{j<i}\exp ({1-\alpha_i})$ is the weight given to a sampled point, $\alpha_i = 1-\exp(-\sigma(\x_i) \delta_i)$ is the opacity and~$\delta_i = t_{i+1} - t_i$.
The MLP parameters are optimized by minimizing the difference between the predicted ray colour and the ground truth ray colour, specifically:
\begin{equation}
    \mathcal{L}(\mathbf{r}; \parsNerf) = \|\mathbf{C}(\mathbf{r}; \parsNerf) - \mathbf{C}_{gt}(\mathbf{r})\|^2
\end{equation}
In this work, we employ the Instant-NGP \cite{mueller2022instant} variant of NeRF for training.

\subsection{Training the quadrature field}
\label{sec:q-field}
The value $w_i$ in \cref{eq:rendering} represents the weight given to the quadrature point $i$; the higher the weight, the more likely light traveling along the direction $d$ hits the point $\mathbf{x}_i$.
In this sense, $w_i$ can be seen as (view-dependent) \textit{surfaceness}, as described by Goli \etal ~\cite{goli2022nerf2nerf}.
While for solid surfaces a single quadrature point should be sufficient to approximate the rendering equation, for non-solid objects more than one quadrature point is needed.
Traditional NeRF models sample a fixed number of quadrature points from the probability density function $w$ to approximate integration via \cref{eq:rendering}. 
However, it is not clear how to determine these quadrature points deterministically. 
In this work, we propose a deterministic way to find quadrature points for \emph{all surface types}.
We emphasize here that we are interested in \emph{multiple} quadrature points along the ray to enable volumetric effects, unlike those that only allow a single point~\cite{yariv2023bakedsdf,chen2022mobilenerf}.

\paragraph{Defining the quadrature field}
We seek to find a field that \textit{concentrates} quadrature points in regions where surfaces are more likely to occur.\footnote{Note that this is not restricted to solids, but also transparent surfaces that require many quadrature points per ray to be accurately represented.}
We take inspiration from parameterization literature~\cite{subdivision2010}, as well as from methods that exploit the zero crossing of a signed distance field to define quadrature points~\cite{Oechsle2021ICCV, yariv2021volume, wang2021neus}, and define our \textit{quadrature field} as:
\begin{align}
\q(\mathbf{x}) &= \sin (\omega \: \f(\mathbf{x}; \parsQuad))
\label{eq:q}
\\
\text{where}& \:\: \f(\mathbf{x}; \parsQuad): \mathbb{R}^3\rightarrow \mathbb{R}
\label{eq:f}
\end{align}
and $\omega$ is a hyperparameter that controls the 
frequency of zero-crossings as shown
\begin{wrapfigure}[11]{r}{0.5\textwidth}
    \includegraphics[width=\linewidth,trim= 0 0 0 30]{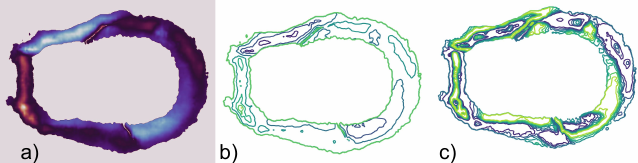}
\caption{\textbf{Effect of omega on quadrature field}. a) quadrature
field along a cross-section of a shape, b) zero-crossings of the quadrature field at $\omega$ = 1 and c) at $\omega$ = 50.
The higher values of omega leads to more zero-crossings.}
    \label{fig:onion}
\end{wrapfigure}
in \cref{fig:onion}. 
Our quadrature points 
are then
defined by the intersection of a ray and the zero crossings of~$\q$.
Note the field $\q$ is only a function of position, as the quadrature points will be represented as a surface mesh, whose geometry does not change according to a viewpoint.
To train the quadrature field, we optimize the parameters~$\parsQuad$ of the function $\f$, and create quadrature points that 
approximate the volume-rendering integral along a given ray.

\paragraph{Training the quadrature field}
For quality rendering,
the field $q$ should have more zero-crossings in the region where the weight function $w$ attains higher values.
To fulfill this objective, we make two simple observations:
\begin{itemize}
\item Assuming local linearity, the number of zero-crossings of~\cref{eq:q} will be proportional to the gradient of~\cref{eq:f};
\item  As the weight function $w$ in \cref{eq:rendering} is a \textit{view-dependent} quantity, we can only supervise the \textit{directional} derivative of \cref{eq:f}.
\end{itemize}
Putting these two observations together, the constraint~$\nabla f(\mathbf{x}; \parsQuad) \cdot \dir \approx w$ emerges, which we approximately satisfy via the following loss:
\begin{equation}
    \mathcal{L}_f(\mathbf{x}; \parsQuad) = ||\nabla \f(\x; \parsQuad) \cdot \mathbf{d}| - \max(w(\x, \dir), w(\x, -\dir))|
    \label{eq:lossf}
\end{equation}
where note the function $w$ is non-optimized and instead kept fixed. 

\begin{wrapfigure}[15]{r}{0.4\textwidth}
    \includegraphics[width=\linewidth, trim = 0 20 0 0]{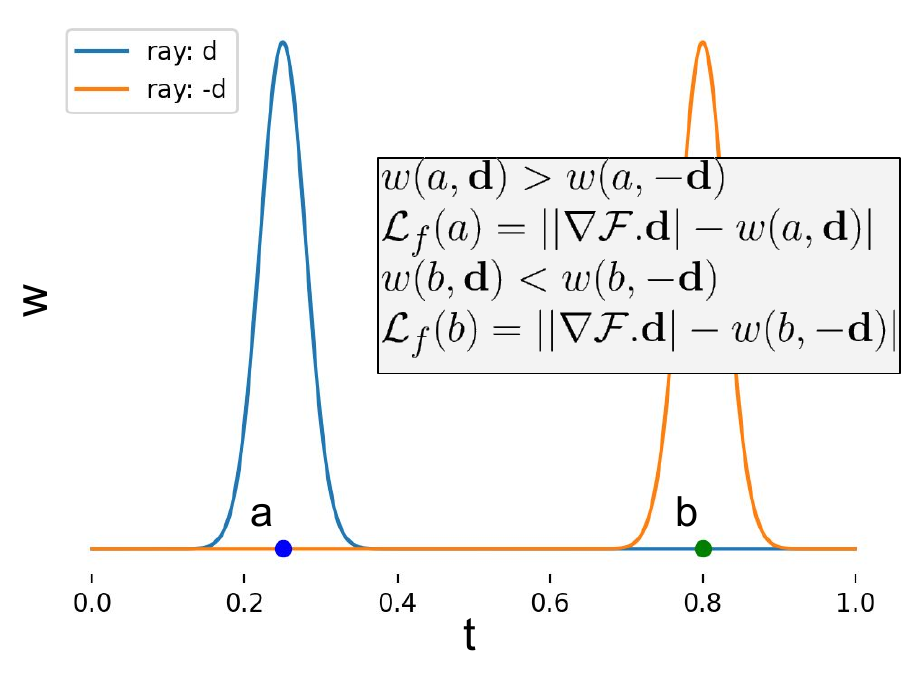}
    \caption{\textbf{Quadrature field loss.} For a particular point, the quad field is supervised to predict the directional gradient to be equal to the maximum weight between the bi-directions (~\cref{eq:lossf}).
    }
    \label{fig:quad-field-loss}
\end{wrapfigure}%
As $w(\mathbf{x}, d)$ is a function of direction, whereas $\nabla f(\mathbf{x})$ is direction independent, we force the field $f$ to vary most in the direction for which $w(\mathbf{x},\pm d)$ is the largest; see~\cref{fig:quad-field-loss}.
We implement $\mathcal{F}$ as an MLP on top of a hash-grid~\cite{mueller2022instant}.

An alternative 
would be to take maximum along all directions as is suggested in nerf2nerf~\cite[Eq.10]{nerf2nerf}, but it would require an impractical amount of compute for training.
MobileNeRF instead initializes a pruning grid to zero in~\cite[Eq.10]{chen2022mobilenerf} and uses surface-field as a lower-bound.
In our case, this entails initializing the network as $\nabla f(\mathbf{\x})=0 \;\forall \x$, which is harder to achieve.
Our formulation, \cref{eq:lossf}, provides simple and stable optimization.

Finally, to encourage \textit{sparse} creation of surfaces, thus fewer quadratures to rasterize, we have explored the use of an additional $\ell_1$ regularization.
We observed that 
the pruning data structure in Instant-NGP~\cite{mueller2022instant} is sufficient to remove surfaces from low-density regions.
After learning the quadrature field,
we extract the quadrature mesh $\mathcal{M}$ via marching cubes on the function~$\q$.

\subsection{Fine-tuning}
\label{sec:finetuning}
While \cref{eq:lossf} can supervise the density of quadrature points, their precise location should be derived by directly optimizing for photometric reconstruction losses.
One way to achieve this would be to fine-tune the vertices of the mesh and the original NeRF directly, as in \cite{chen2022mobilenerf}.
However this saturates GPU memory and, in our experience, leads to non-smooth optimization.
Instead, we employ a vector field to represent deformations continuously in space:
\begin{equation}
\dfield(\x; \parsDef): \mathbb{R}^3 \rightarrow \mathbb{R}^3
\end{equation}
and use a perturbed set of quadrature positions $\{\tilde\x_i\}$ to evaluate photometric reconstruction via \cref{eq:rendering}, where perturbation is restricted to happen \textit{along} the ray direction $\dir$:
\begin{align}
\tilde\x_i = \x_i + \delta(\x_i) \quad \delta(\x_i) = \kappa\:\dir\cdot \dfield(\x_i; \parsDef) \: \dir
\end{align}
where a hyperbolic tangent is used as the final activation function in the $\dfield$, together with $\kappa$, to limit the perturbations within the spatial support of the marching cube mesh and thus stabilize training.
We implement the deformation field using an MLP on top of a hash-grid~\cite{mueller2022instant}. 
Given the deformation field, a mesh vertex $\vertex_i$ is updated as:
\begin{equation}\label{eq:updatevertices}
    \vertex_i \leftarrow \vertex_i + 
    \expect_\iView 
    [w_{i,\iView} (\kappa \dir_\iView \cdot \dfield(\vertex_i; \parsDef)) \: \dir_\iView]
    \: / \:
    \expect_\iView [w_{i,\iView}]
\end{equation}
where $c$ indices over training views, and the perturbation is weighted by the weight $w_{i,\iView}$ on the basis of how much a view $\iView$ affects the given vertex $\vertex_i$.

\paragraph{Training loss and regularization}
Training is done by defining photometric loss with perturbed quadrature points as:
\begin{align}
    \mathcal{L}_\text{def}(\mathbf{r}, \parsNerf, \parsDef) = \|\sum_i w(\tilde\x_i,\mathbf{r})c(\tilde\x_i) - \mathbf{C}_{gt}(\mathbf{r})\|^2
\end{align}
We jointly optimize the parameters of NeRF ($\parsNerf$) and the deformation field ($\parsDef$).
We also encourage $\parsDef$
to be smooth by minimizing the norm of deformation, and by encouraging the deformation to be smooth for each triangle $\triangle$:
\begin{equation}
    \mathcal{L}_\text{reg}(\parsDef) = \expect_{\triangle \in \mesh} \expect_{(\x_a, \x_b) \in \triangle}  \| \dfield(\x_a; \parsDef) \|_2^2  + \|\dfield(\x_a; \parsDef) - \dfield(\x_b; \parsDef)\|_2^2
    \nonumber
\end{equation}

\paragraph{Training implementation}
We employ block coordinate descent, where we first optimize the deformation field till convergence, and then we update the mesh vertices.
In order to update vertices in \cref{eq:updatevertices}, we perform a sweep over all training views to compute $w_i$.
We do not change the topology of the mesh during the above process, as we assume that the extracted mesh from the quadrature field is a good approximation.
\gs{Note that fine-tuning allows our approach to adapt to fewer quadrature points per ray while retaining higher reconstruction quality, as shown in the~\cref{tab:ablation}.}

\subsection{Baking and rendering neural features}
\label{sec:baking}

After fine-tuning, we now prepare the triangular mesh and the texture maps that we ultimately use to render.
We start by post-processing the mesh to remove surfaces that are not visible from training views -- these are likely artifacts as they were never ``seen''.
We further remove the surfaces for which the maximum volumetric weight $w$ across all training views is below a threshold.
We then construct the texture map by first parameterizing the mesh using a publicly available library~\cite{xatlas}.
We provide more detail on parameterization in the supplementary material.

\gs{We implement our real-time rendering using highly optimized Nvidia Optix~\cite{nvidiaoptix} library for ray-tracing on an RTX GPU. For compatibility with Optix ray-tracing pipeline, as well as general efficiency, we compress the representation of colour, alpha, and spherical Gaussian parameters to 8-bit texture maps.
Specifically, we use a sigmoid transformation to bring unbounded RGB coefficients to a $[0, 1]$ range, represent spherical Gaussian lobe axes as 8-bit azimuth and elevation angles, and compress lambda values using a logarithmic mapping.
These quantization results in minimal loss in performance as shown in \cref{tab:ablation}.
}
\subsection{Implementation details}
We implement our NeRF using Instant-NGP, where we use $3-6$ spherical Gaussians depending on the dataset for view-dependent effects.
We use the Nerfacc~\cite{li2023nerfacc} library based on Pytorch~\cite{pytorch}, as it provides stable training at mixed precision with comparable performance as the original Instant-NGP paper.
For real scenes, we use contraction mapping~\cite{barron2022mipnerf360} to train the NeRF.
For extracting meshes, we use a $1024$ sized voxel grid for synthetic scenes and a $2048$ sized voxel grid for real scenes.
We further add meshes extracted from the density field of the NeRF with the mesh extracted from our quadrature field, which helps in avoiding holes like artifacts.
We use $\omega=100$ for synthetic scenes and $\omega=10$ for real scenes.
\gs{We allow maximum of 25 ray-triangle intersection for synthetic scenes and 15 for real scenes.}
We use super-sampling at twice the resolution to achieve anti-aliasing.
We ablate this choice in the \cref{tab:ablation} and provide more information in the Supplementary material.
Our code is available at \url{https://quadraturefields.github.io}.
\section{Experiments}
\label{sec:experiments}
\subsection{Novel-view synthesis}
\begin{figure}[t]
    \centering
    \noindent\includegraphics[width=\columnwidth]{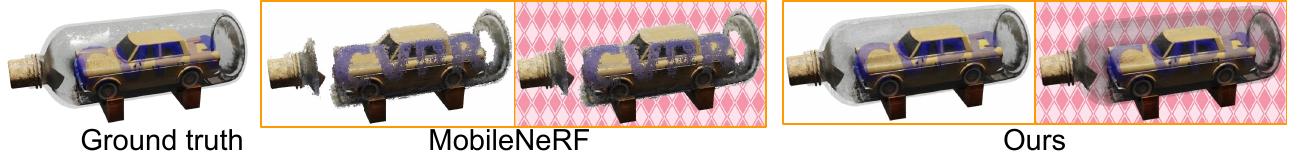} 
    \caption{\gs{Our approach can represent the transparency of a glassy object (Objaverse dataset~\cite{objaverseXL}) achieving 30.6 PSNR
 whereas MobileNeRF fails with only 24.6 PSNR. To effectively test transparency, the shape is also rendered against a textured background. Please zoom-in to see details.
 }}\label{fig:bottled-ship}
\end{figure}
\begin{figure}[t]
    \centering
    \includegraphics[width=\columnwidth]{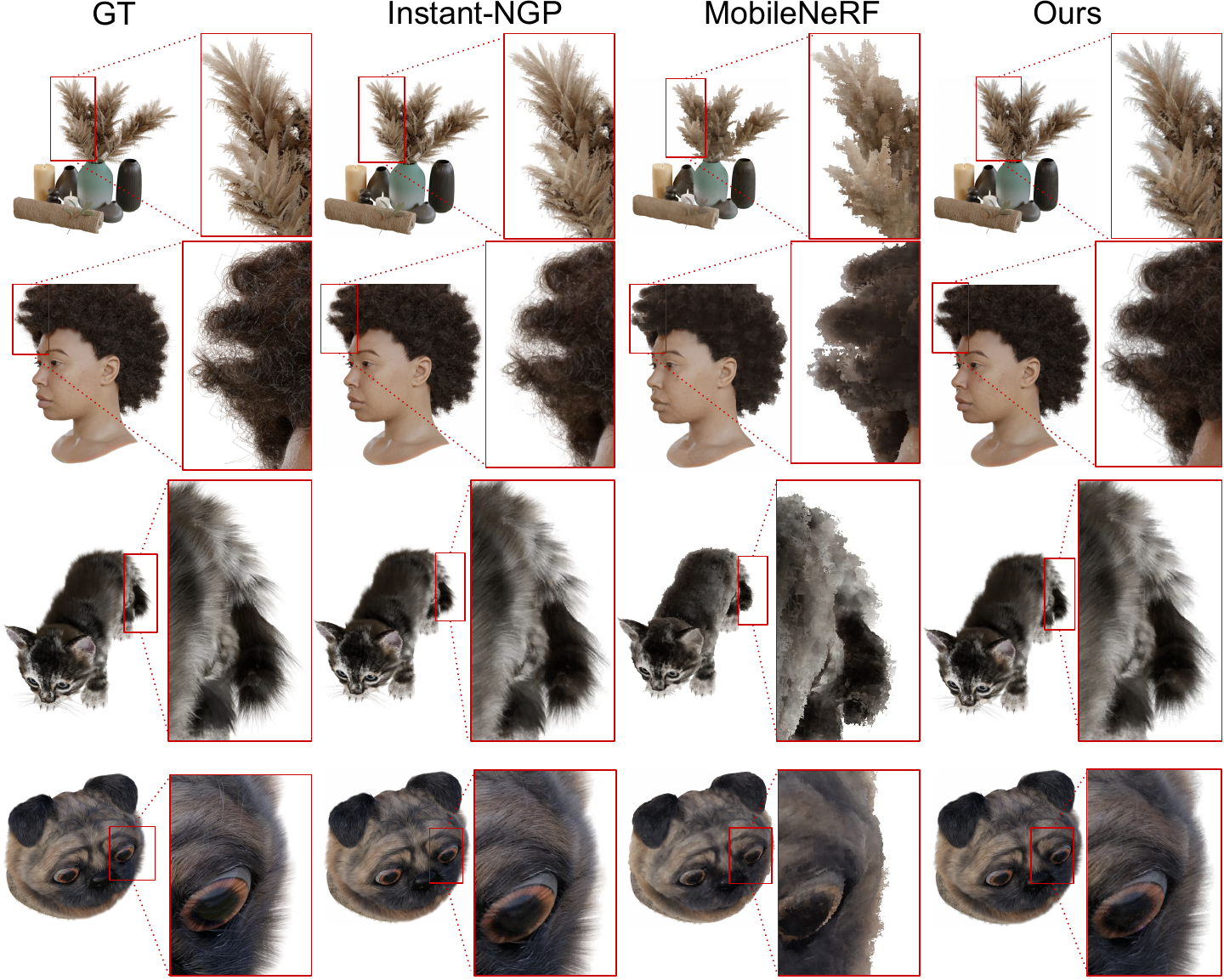}
    \caption{\gs{\textbf{Qualitative results on Shelly-dataset.} Our real-time rendering approach results in a detailed reconstruction of non-opaque objects. Please zoom in to see details.}}
    \label{fig:shelly}
\end{figure}
\begin{table}[t]
\caption{\textbf{Quantitative performance on Shelly dataset~\cite{adaptiveshells2023}.} We compare our approach with various real-time approaches based on baking neural features. 
We also report average number of samples used per-ray to render a pixel.}\label{tab:shelly}
\centering
\resizebox{\textwidth}{!}{%
\setlength{\tabcolsep}{12pt}
\begin{tabular}{@{}lcccccccc@{}}
\toprule
\textbf{Methods}  & \textbf{Avg. \# samples} & \textbf{Avg. PSNR}   & \textbf{Fernvase} & \textbf{Pug}   & \textbf{Woolly} & \textbf{Horse} & \textbf{Khady} & \textbf{Kitten} \\
\hline
Instant-NGP~\cite{mueller2022instant}                &7.13 & 40.13 & 40.99    & 38.75 & 38.66  & 47.05 & 33.00 & 42.31  \\
GS~\cite{kerbl3Dgaussians}                   & - & 38.30 &	39.05	& 38.63	& 34.39	& 45.16	& 31.73	& 40.82\\
\hline
Adaptive shells~\cite{adaptiveshells2023}    & 1.74 & 36.02 & 36.47    & 35.83 & 34.19  & 40.57 & 31.22 & 37.82  \\
Ours (fine-tuned)               & 2.40 & 37.29 & 37.56    & 36.35 & 35.21  & 42.04 & 32.45 & 40.15  \\
\hline
MobileNeRF~\cite{chen2022mobilenerf}         &<1 & 31.62 & 31.38    & 31.50 & 31.61  & 36.48 & 26.84 & 31.93  \\
Ours (baked)     & 2.40 & 35.13 & 35.85    & 34.42 & 31.96  & 38.67 & 31.64 & 38.25 \\
\bottomrule
\end{tabular}%
}
\end{table}
We design our experiments to showcase superior reconstruction quality in real-time on a variety of scenes.
\gs{To showcase the ability of our approach to reconstruct non-opaque shapes we experiment with the Shelly dataset~\cite{adaptiveshells2023} with 6 360-degree scenes with emphasis on fuzzy surfaces with complex geometry and transparency.}
We also use the NeRF-Synthetic dataset consisting of 8 synthetic 360-degree scenes \cite{Mildenhall2020nerf}.
We also experiment with a more challenging real-world dataset consisting of 7 scenes from MipNeRF 360\cite{barron2022mipnerf360}.
We focus on comparing with previous works \gs{that extract meshes to constrain quadrature points used in volumetric rendering. 
We compare with Mobile-NeRF and Baked-SDF that bake neural features into a mesh.}
\gs{We also compare with Adaptive-shells~\cite{adaptiveshells2023} which produces volumetric effects by sampling quadrature points within a shell defined by a mesh.}
\gs{We perform qualitative comparisons with the baselines if their final renderings or code-base are publicly available.}
We evaluate the performance of our method using Peak Signal to Noise Ratio (PSNR). 
Other metrics such as Learned Perceptual Image Patch Similarity (LPIPS) and Structural Similarity Index (SSIM) are provided in the Supp. material.
\gs{We also report the average number of samples per ray to highlight the efficiency of a method in rendering a pixel.}
We provide ablations showing the effect of our design choices in~\cref{tab:ablation}.

\paragraph{Experiments on non-opaque Shapes}
\gs{
We experiment with the Shelly dataset \cite{adaptiveshells2023} introduced by Wang \etal. It covers
a wider variety of appearances including fuzzy surfaces such as hair,
fur, and foliage rendered at $1920 \times 1080$ resolution.
Our approach faithfully reconstructs thin structures and transparency as shown in~\cref{fig:shelly}.
Though our approach loses performance because of the baking process, it still produces quantitatively similar results to Adaptive shells as shown in~\cref{tab:shelly}.
To further stress test our approach, we experiment with a transparent object from the Objaverse dataset~\cite{objaverseXL} in the ~\cref{fig:bottled-ship}.
MobileNeRF fails to reconstruct thin structures and completely fails to reconstruct the transparent shape, shown in the~\cref{fig:shelly} and~\cref{fig:bottled-ship}.
}

\paragraph{NeRF synthetic dataset}
\begin{table}[t]
\caption{\textbf{Quantitative performance on NeRF synthetic dataset.} We compare our approach with various real-time approaches that are based on baking neural features using PSNR metric. We also report average \# samples to render a pixel.}
\label{tab:nerf-synthetic}
\centering
\resizebox{\linewidth}{!}{%
\setlength{\tabcolsep}{12pt}
\begin{tabular}{@{}lclcccccccc@{}} 
\toprule
\textbf{Methods}                      & \textbf{Avg. \# samples} & \textbf{Avg. PSNR}  & \textbf{Lego}  & \textbf{Chair} & \textbf{Ship}  & \textbf{Mic}   & \textbf{Drums} & \textbf{Mat.} & \textbf{Ficus} & \textbf{Hotd.}\\
\hline
Instant-NGP~\cite{mueller2022instant} &7.95	&	33.16 &	35.84 &	35.54 &	30.71 &	36.58 &	25.60 &	29.58 &	33.98 &	37.48 \\
GS~\cite{kerbl3Dgaussians} & - &	33.31 &	35.78 &	35.83 &	30.80 &	35.36 &	26.15 &	30.00 &	34.87 &	37.72 \\
\hline
SNeRG  \cite{snerg2021}       &- &	30.38 &	33.82 & 33.24 &	27.97 &	32.60 &	24.57 &	27.21 &	29.32 &	34.33\\
VMesh \cite{guo2023vmesh}           & -&	30.70 & - & - & - & - & - & - & - & -\\
MobileNeRF  \cite{chen2022mobilenerf}  & <1&	30.90 \brzc & 34.18 & 34.09 & 29.06 & 32.48 & 25.02 & 26.72 & 30.20 & 35.46\\
Adaptive-shells~\cite{adaptiveshells2023} &	3.53 &	31.84 \gc &	33.49 &	34.94 &	29.54 &	33.91 &	25.19 &	27.82 &	33.63 &	36.21 \\
Ours & 1.69	&	31.00 \silc &	32.89 &	33.48 &	28.83 &	33.70 &	25.33 &	27.91 &	32.24 &	33.57 \\
\bottomrule
\end{tabular}%
}
\end{table}
Our work produces triangular meshes using the quadrature field that is based on the surface field (\cref{sec:q-field}).
An alternative approach to generate meshes would be to use Delaunay Triangulation on the surface field.
This approach produces more vertices of tetrahedra in regions where the surface is likely to occur.
The mesh reconstructed using Delaunay triangulation results in a bad representation of the underlying surface as is shown in the Supplementary material.
We further compare with other methods that propose baking neural features into a volumetric representation such as voxels (SNeRG~\cite{snerg2021}), and meshes (MobileNeRF~\cite{chen2022mobilenerf}) and hybrid volumetric and mesh representation (VMesh~\cite{guo2023vmesh}).
We report the results on the NeRF synthetic dataset in the~\cref{tab:nerf-synthetic}.

\paragraph{MipNeRF 360 dataset}
\begin{table*}[t]
\caption{\textbf{Qualitative evaluation on mip-NeRF 360 dataset.}  \gs{We report reconstruction quality (PSNR metric). We also report average \# samples to render a pixel.}
}
\label{tab:mipnerf360}
\centering
\resizebox{\textwidth}{!}{
\setlength{\tabcolsep}{10pt}
\begin{tabular}{@{}lclllllllll@{}}
\toprule
\textbf{Method}       & \textbf{Avg.\# samples}   & \textbf{Mean-indoor}& \textbf{Kitchen}  & \textbf{Room}  & \textbf{Bonsai} & \textbf{Counter} & \textbf{Mean-outdoor} & \textbf{Garden} & \textbf{Bicycle} & \textbf{Stump} \\
\hline
Instant-NGP~\cite{mueller2022instant}	&13.03 & 29.67 & 	29.67 &	30.60 &	31.14 &	27.26 &	24.58 &	25.68 &	23.32 &	24.75  \\
GS~\cite{kerbl3Dgaussians} &- &	30.41 &	30.32 &	30.63 &	31.98 &	28.70 &	26.40 &	27.41 &	25.25 &	26.55 \\
\hline
MobileNeRF~\cite{chen2022mobilenerf}      & <1&	-     &	-     &	-     &	-     &	-     &	23.06         &	23.54 &	21.70 &	23.95 \\
BakedSDF~\cite{yariv2023bakedsdf}         & <1&	27.06 \brzc &	26.72 &	28.68 &	27.17 &	25.69 &	23.52 \silc         &	24.94 &	22.04 &	23.59 \\
Adaptive Shells~\cite{adaptiveshells2023} & 17.41 &	29.19 \gc &	28.43 &	30.63 &	32.47 &	25.24 &	23.17 \brzc     &	25.35 &	22.19 &	21.96 \\
Ours                                      & 5.15 &	28.13 \silc &	28.52 &	28.81 &	28.87 &	26.30 &	24.12 \gc  &	25.54 &	22.93 &	23.90 \\
\bottomrule
\end{tabular}
}
\end{table*}

\begin{figure*}
    \centering
    \includegraphics[width=\linewidth]{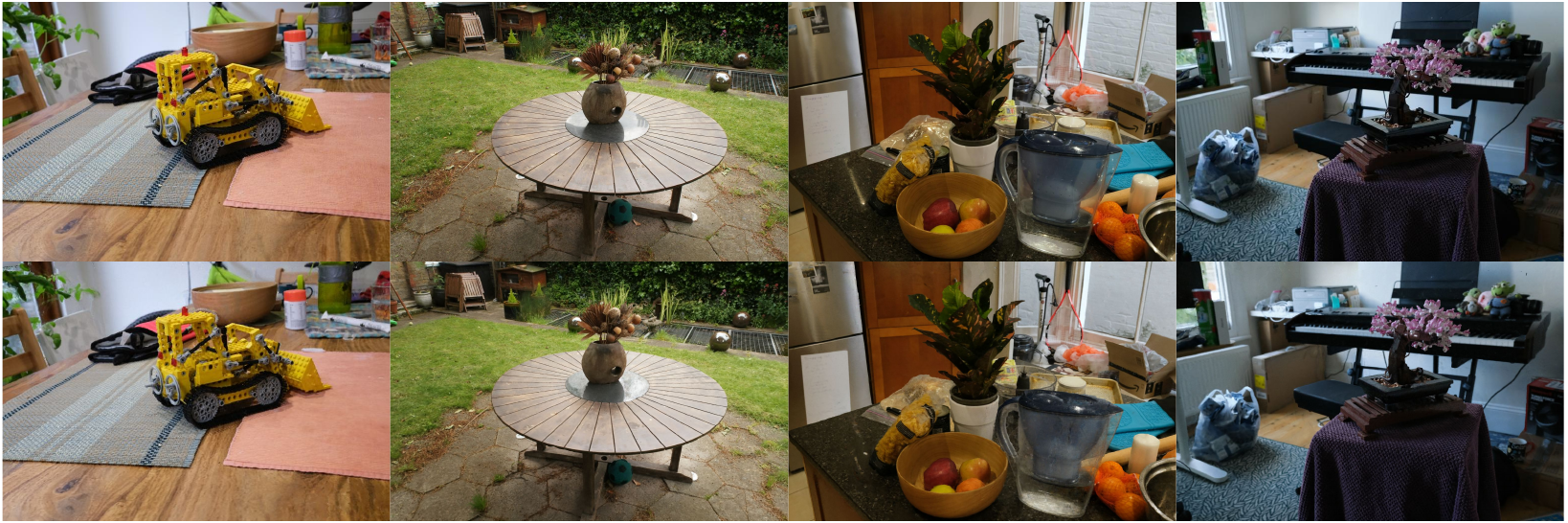} 
    \noindent\caption{\textbf{Visualization of Mip-NeRF 360 dataset.} \textbf{Top:} GT, \textbf{Bottom:} Ours.
    }\label{fig:mipnerf-results}
\end{figure*}

\begin{figure}[t]
    \centering
    \noindent\includegraphics[width=\linewidth]{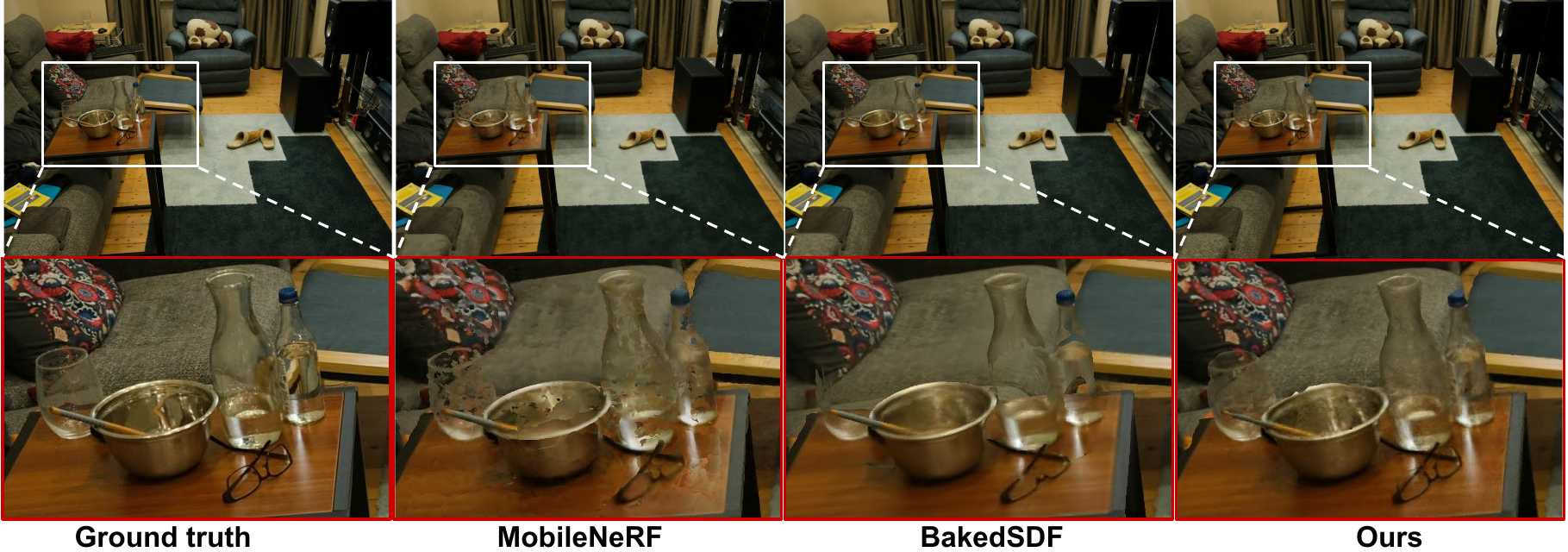} 
    \caption{\textbf{Comparison with BakedSDF~\cite{yariv2023bakedsdf} and MobileNeRF~\cite{chen2022mobilenerf}.} The bottom row shows a zoomed-in reconstruction of transparent objects.}
    \label{fig:comparison-with-bakedsdf}
\end{figure}

\gs{We further evaluate our approach with a real-world dataset and compare it with surface-based representation (MobileNeRF and BakedSDF)
and hybrid volumetric rendering (Adaptive-shells~\cite{adaptiveshells2023}).}
\cref{tab:mipnerf360} shows the quantitative performance of our approach and comparison with the baselines.
\cref{fig:mipnerf-results} shows the qualitative performance of our approach for different real scenes.
Our approach outperforms the surface-based baselines in outdoor scenes and is competitive with the baselines in indoor scenes.
Our approach performs similarly to Adaptive-shells while using only a third of the samples to render images.
\cref{fig:comparison-with-bakedsdf} shows a comparison with surface-based methods --  BakedSDF and MobileNeRF produce incomplete reconstruction of transparent objects.
\subsection{Run-time performance}
\begin{table}[t]
\caption{
\textbf{Run-time comparison across different datasets.} We report our run-time at 1 and 2$\times$ super-sampling.}\label{tab:fps}
\centering
\resizebox{\linewidth}{!}{%
\setlength{\tabcolsep}{18pt}
\begin{tabular}{@{}l|l|cc|cc|cc@{}}
\hline
                                                           &           & \multicolumn{2}{c|}{\textbf{NeRF synthetic}} & \multicolumn{2}{c|}{\textbf{Shelly}} & \multicolumn{2}{c}{\textbf{MipNeRF 360}} \\ \hline
Method                                                     & Framework & FPS             & Watts            & FPS         & Watts        & FPS           & Watts           \\ \hline
GS           & CUDA                       & 393                & 340                & 403  &  340     & 134 & 300   \\
BakedSDF                                                   & WebGL                      & -                &     -             & -   & -     & 72  & 85    \\
MobileNeRF & WebGL                      & 744.5             & 250              & 455   & 340     & 280 & 250   \\
Adaptive Shells                                            & CUDA                       & 281              & 450              & 263 & 450   & 36  & 450   \\
\hline
Ours 1$\times$ (Nvidia 3090)                               & CUDA                       & 570              & 340              & 323& 340   & 140 & 340   \\
Ours 2$\times$ (Nvidia 3090)                               & CUDA                       & 167              & 340              & 93  & 340   & 35  & 340   \\ \hline
\end{tabular}%
}
\end{table}%
We evaluate our run-time performance using the CUDA Optix library running on 
an Nvidia RTX 3090 GPU in the \cref{tab:fps}.
We render the images at $800 \times 800$ resolution for NeRF synthetic dataset and $1920 \times 1080$ for MipNeRF and Shelly dataset.
Our approach runs at an interactive speed or better across all datasets despite using more than one samples per ray.
\gs{We also implement rendering using the depth peeling algorithm~\cite{depthpeeling} to efficiently render in a browser.
These experiments are discussed in the Supplementary material.
}
\subsection{Ablations}
We ablate our algorithms in \cref{tab:ablation} and observe the following:

\begin{itemize}
    \item \textbf{Number of spherical-gaussian lobes:} Increasing spherical lobes improves reconstruction of view-dependent effects, albeit at the cost of rendering.
    
    \item \textbf{Include mesh from density field:} Often including coarse mesh extracted from the density fields complements the mesh extracted from the quadrature fields and helps fill the holes. We provide visualization in the Supp. material.
    
    \item \textbf{Effect of omega:} Larger omega leads to more quadrature points, which better captures the volumetric effects.
    
    \item \textbf{Effect of fine-tuning:} Fine-tuning aligns the mesh with the NeRF density field improving the reconstruction.
    
    \item \textbf{Size of texture map:} Increasing texels per triangle improves reconstruction but at \gs{a slight} cost of speed.
    
\item \textbf{Size of mesh:} A higher resolution mesh gives better reconstruction quality at the cost of rendering speed.

\item \textbf{Super-sampling:} We can achieve anti-aliasing using super-sampling of each pixel resulting in better reconstruction quality at the cost of rendering speed.

\end{itemize}

\begin{table}[t]
\caption{\textbf{Ablations.} \gs{We ablate our proposed approach at three stages using 1) a different number of spherical Gaussian (SG) lobes, 2) whether we use density mesh along with mesh extracted from the quadrature field, 3) the effect of fine-tuning, 4) the use of continuous rendering loss, 5) effect of $\omega$ used for mesh extraction, 6) effect of quantization (baking), 7) size of the mesh, 9) finally the different sizes of texture map used for baking neural features and the effect of super-sampling on reconstruction quality and speed. We use the Mic scene from the synthetic dataset to ablate.}}
\centering
\resizebox{\textwidth}{!}{%
\setlength{\tabcolsep}{10pt}
\begin{tabular}{@{}lllllllll@{}}
\toprule
                                         &                                  &                          & \textbf{PSNR}                       &                                           &                                 &                         & \textbf{PSNR}  & \textbf{FPS} \\ \hline
\multicolumn{1}{l|}{I}                   &           Full model                       & Instant-NGP                      & \multicolumn{1}{l|}{36.58} & \multicolumn{1}{l|}{\multirow{8}{*}{III}} & \multirow{2}{*}{Quantization}   & Ours w/o quant.         & 34.06 & -   \\ \cline{1-4}
\multicolumn{1}{l|}{\multirow{7}{*}{II}} & Finetune                         & No finetune              & \multicolumn{1}{l|}{31.05} & \multicolumn{1}{l|}{}                     &                                 & Ours w/ quant. (baked)          & 33.70 & 210 \\ \cline{2-4} \cline{6-9} 
\multicolumn{1}{l|}{}                    & \multirow{2}{*}{SG lobes}        & Finetune (w/ 3 SG)       & \multicolumn{1}{l|}{33.68} & \multicolumn{1}{l|}{}                     & \multirow{2}{*}{Mesh size}      & Mesh w/ 880k faces      & 33.70 & 210 \\
\multicolumn{1}{l|}{}                    &                                  & Finetune (w/ 6 SG)       & \multicolumn{1}{l|}{34.06} & \multicolumn{1}{l|}{}                     &                                 & Mesh w/ 3.3 mil faces   & 33.92 & 180 \\ \cline{2-4} \cline{6-9} 
\multicolumn{1}{l|}{}                    & \multirow{3}{*}{Mesh Extraction} & Finetune w/ density mesh & \multicolumn{1}{l|}{30.37} & \multicolumn{1}{l|}{}                     & \multirow{2}{*}{Texture size}   & Baked w/ 4096 text. map & 33.11 & 230 \\
\multicolumn{1}{l|}{}                    &                                  & Finetune with quad mesh  & \multicolumn{1}{l|}{33.40} & \multicolumn{1}{l|}{}                     &                                 & Baked w/ 8192 text. map & 33.70 & 210 \\ \cline{6-9} 
\multicolumn{1}{l|}{}                    &                                  & Finetune with omega=10   & \multicolumn{1}{l|}{31.09} & \multicolumn{1}{l|}{}                     & \multirow{2}{*}{Super-sampling} & 1x                      & 32.05 & 750 \\ \cline{2-4}
\multicolumn{1}{l|}{}                    & Loss                             & Finetune w/o cont. loss  & \multicolumn{1}{l|}{34.00} & \multicolumn{1}{l|}{}                     &                                 & 2x                      & 33.70 & 210 \\
\bottomrule
\end{tabular}
\label{tab:ablation}
}
\end{table}

\section{Conclusion}
Our research addresses a critical limitation of the NeRF representation by introducing a novel approach that leverages
textured polygons with continuous opacity and encodes feature vectors, enabling rapid rendering and integration into standard graphics pipelines. 
By training a specialized field to identify quadrature points and utilizing a novel gradient-based loss function, we achieve a quality mesh suitable for interactive rendering on desktops. 
Our method retains the ability to handle scenes featuring transparent objects, enhancing its practical applicability and potential impact in computer vision and graphics domains.

\paragraph{Limitations} 
Our method is bound by the limitations of NeRF; extending quadrature fields to more general rendering techniques would be interesting.
Dealing with thin surfaces poses a challenge, as rarely, our method may miss thin surfaces when limited in capacity; recent developments for better quadrature for NeRFs~\cite{uy-plnerf-neurips23} might be helpful.
For large-scenes, reducing the memory footprint could be interesting to enable extremely low-end devices.

\section*{Acknowledgements}
The authors would like to thank Sara Sabour for helping to create a dataset. This work was supported in part by the Natural Sciences and Engineering Research Council of Canada (NSERC) Discovery Grant, NSERC Collaborative Research and Development Grant, Google, Digital Research Alliance of Canada, and Advanced Research Computing at the University of British Columbia.


%
%
\bibliographystyle{splncs04}
\bibliography{main}

\begin{thebibliography}{10}
\providecommand{\url}[1]{\texttt{#1}}
\providecommand{\urlprefix}{URL }
\providecommand{\doi}[1]{https://doi.org/#1}

\bibitem{attal2022learning}
Attal, B., Huang, J.B., Zollh{\"o}fer, M., Kopf, J., Kim, C.: Learning neural
  light fields with ray-space embedding networks. In: Proceedings of the
  IEEE/CVF Conference on Computer Vision and Pattern Recognition (CVPR) (2022)

\bibitem{attal6d0f}
Attal, B., Ling, S., Gokaslan, A., Richardt, C., Tompkin, J.: {MatryODShka}:
  Real-time {6DoF} video view synthesis using multi-sphere images. In: European
  Conference on Computer Vision (ECCV) (Aug 2020),
  \url{https://visual.cs.brown.edu/matryodshka}

\bibitem{barron2022mipnerf360}
Barron, J.T., Mildenhall, B., Verbin, D., Srinivasan, P.P., Hedman, P.:
  Mip-nerf 360: Unbounded anti-aliased neural radiance fields. CVPR  (2022)

\bibitem{OmniPhotos}
Bertel, T., Yuan, M., Lindroos, R., Richardt, C.: Omniphotos: Casual 360° vr
  photography. ACM Trans. Graph.  (2020)

\bibitem{lumigraph2023}
Buehler, C., Bosse, M., McMillan, L., Gortler, S., Cohen, M.: Unstructured
  Lumigraph Rendering (2023)

\bibitem{Chen2022ECCV}
Chen, A., Xu, Z., Geiger, A., Yu, J., Su, H.: Tensorf: Tensorial radiance
  fields. In: European Conference on Computer Vision (ECCV) (2022)

\bibitem{chen2022mobilenerf}
Chen, Z., Funkhouser, T., Hedman, P., Tagliasacchi, A.: Mobilenerf: Exploiting
  the polygon rasterization pipeline for efficient neural field rendering on
  mobile architectures. arXiv preprint arXiv:2208.00277  (2022)

\bibitem{elu2015}
Clevert, D.A., Unterthiner, T., Hochreiter, S.: Fast and accurate deep network
  learning by exponential linear units (elus). Under Review of ICLR2016 (1997)
  (11 2015)

\bibitem{objaverseXL}
Deitke, M., Liu, R., Wallingford, M., Ngo, H., Michel, O., Kusupati, A., Fan,
  A., Laforte, C., Voleti, V., Gadre, S.Y., VanderBilt, E., Kembhavi, A.,
  Vondrick, C., Gkioxari, G., Ehsani, K., Schmidt, L., Farhadi, A.:
  Objaverse-xl: A universe of 10m+ 3d objects. arXiv preprint arXiv:2307.05663
  (2023)

\bibitem{depthpeeling}
Everitt, C.: Interactive order-independent transparency (2001),
  \url{https://www.gamedevs.org/uploads/interactive-order-independent-transparency.pdf},
  accessed on November 12, 2023

\bibitem{Flynn2019}
Flynn, J., Broxton, M., Debevec, P., DuVall, M., Fyffe, G., Overbeck, R.,
  Snavely, N., Tucker, R.: Deepview: View synthesis with learned gradient
  descent. In: 2019 IEEE/CVF Conference on Computer Vision and Pattern
  Recognition (CVPR). pp. 2362--2371 (2019). \doi{10.1109/CVPR.2019.00247}

\bibitem{yu_and_fridovichkeil2021plenoxels}
{Fridovich-Keil and Yu}, Tancik, M., Chen, Q., Recht, B., Kanazawa, A.:
  Plenoxels: Radiance fields without neural networks. In: CVPR (2022)

\bibitem{Furukawa2015}
Furukawa, Y., Hern\'{a}ndez, C.: Multi-view stereo: A tutorial. Found. Trends.
  Comput. Graph. Vis. p. 1–148 (jun 2015)

\bibitem{gao2020deftet}
Gao, J., Chen, W., Xiang, T., Tsang, C.F., Jacobson, A., McGuire, M., Fidler,
  S.: Learning deformable tetrahedral meshes for 3d reconstruction. In:
  Advances In Neural Information Processing Systems (2020)

\bibitem{Garbin2021FastNeRFHN}
Garbin, S.J., Kowalski, M., Johnson, M., Shotton, J., Valentin, J.P.C.:
  Fastnerf: High-fidelity neural rendering at 200fps. 2021 IEEE/CVF
  International Conference on Computer Vision (ICCV) pp. 14326--14335 (2021),
  \url{https://api.semanticscholar.org/CorpusID:232270138}

\bibitem{goli2022nerf2nerf}
Goli, L., Rebain, D., Sabour, S., Garg, A., Tagliasacchi, A.: {nerf2nerf}:
  Pairwise registration of neural radiance fields  (2022)

\bibitem{nerf2nerf}
Goli, L., Rebain, D., Sabour, S., Garg, A., Tagliasacchi, A.: {nerf2nerf}:
  Pairwise registration of neural radiance fields. In: International Conference
  on Robotics and Automation (ICRA) (2022)

\bibitem{lumigraph}
Gortler, S.J., Grzeszczuk, R., Szeliski, R., Cohen, M.F.: The lumigraph.
  SIGGRAPH '96 (1996)

\bibitem{guo2023vmesh}
Guo, Y.C., Cao, Y.P., Wang, C., He, Y., Shan, Y., Qie, X., Zhang, S.H.: Vmesh:
  Hybrid volume-mesh representation for efficient view synthesis (2023)

\bibitem{subdivision2010}
He, L., Schaefer, S., Hormann, K.: Parameterizing subdivision surfaces. ACM
  Trans. Graph.  (2010)

\bibitem{snerg2021}
Hedman, P., Srinivasan, P.P., Mildenhall, B., Barron, J.T., Debevec, P.: Baking
  neural radiance fields for real-time view synthesis (2021)

\bibitem{freeviewhedman2018}
Hedman, P., Philip, J., Price, T., Frahm, J.M., Drettakis, G., Brostow, G.:
  Deep blending for free-viewpoint image-based rendering. ACM Trans. Graph.
  (2018)

\bibitem{ReluField_sigg_22}
Karnewar, A., Ritschel, T., Wang, O., Mitra, N.: Relu fields: The little
  non-linearity that could. In: ACM SIGGRAPH 2022 Conference Proceedings.
  SIGGRAPH '22, Association for Computing Machinery, New York, NY, USA (2022).
  \doi{10.1145/3528233.3530707}, \url{https://doi.org/10.1145/3528233.3530707}

\bibitem{psr2006}
Kazhdan, M., Bolitho, M., Hoppe, H.: {Poisson Surface Reconstruction}. In:
  Sheffer, A., Polthier, K. (eds.) Symposium on Geometry Processing. The
  Eurographics Association (2006). \doi{10.2312/SGP/SGP06/061-070}

\bibitem{kerbl3Dgaussians}
Kerbl, B., Kopanas, G., Leimk{\"u}hler, T., Drettakis, G.: 3d gaussian
  splatting for real-time radiance field rendering. ACM Transactions on
  Graphics  \textbf{42}(4) (July 2023),
  \url{https://repo-sam.inria.fr/fungraph/3d-gaussian-splatting/}

\bibitem{Levoy96}
Levoy, M., Hanrahan, P.: Light field rendering. In: Proceedings of the 23rd
  Annual Conference on Computer Graphics and Interactive Techniques. SIGGRAPH
  '96 (1996)

\bibitem{li2023nerfacc}
Li, R., Gao, H., Tancik, M., Kanazawa, A.: Nerfacc: Efficient sampling
  accelerates nerfs. arXiv preprint arXiv:2305.04966  (2023)

\bibitem{marchingcubes}
Lorensen, W.E., Cline, H.E.: Marching cubes: A high resolution 3d surface
  construction algorithm. SIGGRAPH Comput. Graph. p. 163–169 (aug 1987)

\bibitem{mildenhall2019llff}
Mildenhall, B., Srinivasan, P.P., Ortiz-Cayon, R., Kalantari, N.K.,
  Ramamoorthi, R., Ng, R., Kar, A.: Local light field fusion: Practical view
  synthesis with prescriptive sampling guidelines. ACM Transactions on Graphics
  (TOG)  (2019)

\bibitem{Mildenhall2020nerf}
Mildenhall, B., Srinivasan, P.P., Tancik, M., Ramamoorthi, R., Ng, R.: Nerf:
  Representing scenes as neural radiance fields for view synthesis. In: ECCV
  (2020)

\bibitem{mueller2022instant}
M\"uller, T., Evans, A., Schied, C., Keller, A.: Instant neural graphics
  primitives with a multiresolution hash encoding. ACM TOG  (2022)

\bibitem{Munkberg_2022_CVPR}
Munkberg, J., Hasselgren, J., Shen, T., Gao, J., Chen, W., Evans, A., M\"uller,
  T., Fidler, S.: {Extracting Triangular 3D Models, Materials, and Lighting
  From Images}. In: Proceedings of the IEEE/CVF Conference on Computer Vision
  and Pattern Recognition (CVPR). pp. 8280--8290 (June 2022)

\bibitem{Oechsle2021ICCV}
Oechsle, M., Peng, S., Geiger, A.: Unisurf: Unifying neural implicit surfaces
  and radiance fields for multi-view reconstruction. In: International
  Conference on Computer Vision (ICCV) (2021)

\bibitem{Park_2019_CVPR}
Park, J.J., Florence, P., Straub, J., Newcombe, R., Lovegrove, S.: Deepsdf:
  Learning continuous signed distance functions for shape representation. In:
  The IEEE Conference on Computer Vision and Pattern Recognition (CVPR) (June
  2019)

\bibitem{nvidiaoptix}
Parker, S.G., Bigler, J., Dietrich, A., Friedrich, H., Hoberock, J., Luebke,
  D., McAllister, D., McGuire, M., Morley, K., Robison, A., Stich, M.: Optix: a
  general purpose ray tracing engine. ACM Trans. Graph.  \textbf{29}(4) (jul
  2010). \doi{10.1145/1778765.1778803},
  \url{https://doi.org/10.1145/1778765.1778803}

\bibitem{pytorch}
Paszke, A., Gross, S., Massa, F., Lerer, A., Bradbury, J., Chanan, G., Killeen,
  T., Lin, Z., Gimelshein, N., Antiga, L., Desmaison, A., Kopf, A., Yang, E.,
  DeVito, Z., Raison, M., Tejani, A., Chilamkurthy, S., Steiner, B., Fang, L.,
  Bai, J., Chintala, S.: Pytorch: An imperative style, high-performance deep
  learning library. In: Advances in Neural Information Processing Systems 32
  (2019),
  \url{http://papers.neurips.cc/paper/9015-pytorch-an-imperative-style-high-performance-deep-learning-library.pdf}

\bibitem{Soft3DReconstruction}
Penner, E., Zhang, L.: Soft 3d reconstruction for view synthesis
  \textbf{36}(6) (2017)

\bibitem{derf}
Rebain, D., Jiang, W., Yazdani, S., Li, K., Yi, K.M., Tagliasacchi, A.: Derf:
  Decomposed radiance fields. In: 2021 IEEE/CVF Conference on Computer Vision
  and Pattern Recognition (CVPR). pp. 14148--14156 (2021).
  \doi{10.1109/CVPR46437.2021.01393}

\bibitem{kilonerf}
Reiser, C., Peng, S., Liao, Y., Geiger, A.: Kilonerf: Speeding up neural
  radiance fields with thousands of tiny mlps. In: International Conference on
  Computer Vision (ICCV) (2021)

\bibitem{reiser2023merf}
Reiser, C., Szeliski, R., Verbin, D., Srinivasan, P.P., Mildenhall, B., Geiger,
  A., Barron, J.T., Hedman, P.: Merf: Memory-efficient radiance fields for
  real-time view synthesis in unbounded scenes. arXiv preprint arXiv:2302.12249
   (2023)

\bibitem{rematas2022urf}
Rematas, K., Liu, A., Srinivasan, P.P., Barron, J.T., Tagliasacchi, A.,
  Funkhouser, T., Ferrari, V.: Urban radiance fields. CVPR  (2022)

\bibitem{Riegler2020FVS}
Riegler, G., Koltun, V.: Free view synthesis. In: European Conference on
  Computer Vision (2020)

\bibitem{Riegler2021SVS}
Riegler, G., Koltun, V.: Stable view synthesis. In: Proceedings of the IEEE
  Conference on Computer Vision and Pattern Recognition (2021)

\bibitem{shen2021dmtet}
Shen, T., Gao, J., Yin, K., Liu, M.Y., Fidler, S.: Deep marching tetrahedra: a
  hybrid representation for high-resolution 3d shape synthesis. In: Advances in
  Neural Information Processing Systems (NeurIPS) (2021)

\bibitem{SOTArendering}
Tewari, A., Fried, O., Thies, J., Sitzmann, V., Lombardi, S., Sunkavalli, K.,
  Martin~Brualla, R., Simon, T., Saragih, J., Nießner, M., Pandey, R.,
  Fanello, S., Wetzstein, G., Zhu, J.Y., Theobalt, C., Agrawala, M., Shechtman,
  E., Goldman, D., Zollhöfer, M.: State of the art on neural rendering.
  Computer Graphics Forum  (2020)

\bibitem{uy-plnerf-neurips23}
Uy, M.A., Nakayama, G.K., Yang, G., Thomas, R.K., Guibas, L., Li, K.: Nerf
  revisited: Fixing quadrature instability in volume rendering. In: NeurIPS
  (2023)

\bibitem{verbin2022refnerf}
Verbin, D., Hedman, P., Mildenhall, B., Zickler, T., Barron, J.T., Srinivasan,
  P.P.: {Ref-NeRF}: Structured view-dependent appearance for neural radiance
  fields. CVPR  (2022)

\bibitem{wang2023nemto}
Wang, D., Zhang, T., S{\"u}sstrunk, S.: {NEMTO: Neural Environment Matting for
  Novel View and Relighting Synthesis of Transparent Objects}. In: Proceedings
  of the IEEE/CVF International Conference on Computer Vision (ICCV) (October
  2023)

\bibitem{wang2021neus}
Wang, P., Liu, L., Liu, Y., Theobalt, C., Komura, T., Wang, W.: Neus: Learning
  neural implicit surfaces by volume rendering for multi-view reconstruction.
  arXiv preprint arXiv:2106.10689  (2021)

\bibitem{adaptiveshells2023}
Wang, Z., Shen, T., Nimier-David, M., Sharp, N., Gao, J., Keller, A., Fidler,
  S., M\"uller, T., Gojcic, Z.: Adaptive shells for efficient neural radiance
  field rendering. ACM Trans. Graph.  \textbf{42}(6) (2023).
  \doi{10.1145/3618390}, \url{https://doi.org/10.1145/3618390}

\bibitem{weiping2018autoint}
Weiping, S., Chence, S., Zhiping, X., Zhijian, D., Yewen, X., Ming, Z., Jian,
  T.: Autoint: Automatic feature interaction learning via self-attentive neural
  networks. arXiv preprint arXiv:1810.11921  (2018)

\bibitem{wu2023alphasurf}
Wu, T., Liang, H., Zhong, F., Riegler, G., Vainer, S., Oztireli, C.:
  $\alpha$surf: Implicit surface reconstruction for semi-transparent and thin
  objects with decoupled geometry and opacity (2023)

\bibitem{yariv2021volume}
Yariv, L., Gu, J., Kasten, Y., Lipman, Y.: Volume rendering of neural implicit
  surfaces. In: Thirty-Fifth Conference on Neural Information Processing
  Systems (2021)

\bibitem{yariv2023bakedsdf}
Yariv, L., Hedman, P., Reiser, C., Verbin, D., Srinivasan, P.P., Szeliski, R.,
  Barron, J.T., Mildenhall, B.: Bakedsdf: Meshing neural sdfs for real-time
  view synthesis. arXiv  (2023)

\bibitem{xatlas}
Young, J.: Xatlas. Available at: \url{https://github.com/jpcy/xatlas} (2023)

\bibitem{yu2021plenoctrees}
Yu, A., Li, R., Tancik, M., Li, H., Ng, R., Kanazawa, A.: {PlenOctrees} for
  real-time rendering of neural radiance fields. In: ICCV (2021)

\end{thebibliography}
\clearpage
\section{Supplementary Material}
In this section, we provide information about the following:
\begin{itemize}
    \item Surface parameterization
    \item Spherical Gaussian fitting
    \item Run-time performance on mipNeRF 360 scenes
    \item Visualization of ablations (quad mesh)
    \item Visualization of mesh extracted using different approaches
    \item Evaluation using more metrics
    \item Storage requirements of rendering
    \item Visualization of the geometry of scenes
\end{itemize}

\subsection{Spherical Gaussian fitting}
To achieve high rendering speed, we parameterize radiance using Spherical Gaussians (SG).
We start with a fine-tuned NeRF (see Sec. 3.3) in which view-dependent radiance is represented using an MLP.
We then train an instant-NGP back-bone to predict radiance on the surface of the mesh using a fixed number of SG.
This network is supervised using rendering loss (Eq. 2).
We use 6 SG lobes for NeRF synthetic dataset that consists of shapes with large view-dependent effects.
We use 3 SG lobes for Shelly dataset and mipNeRF 360 scenes, which provides a good tradeoff between rendering speed and reconstruction quality.
This trained network is used to bake SG parameters into a texture map (see Sec 3.4).
We choose this two-step process instead of directly using SGs to represent radiance in the pre-training of NeRF because smaller number of SG lobes have less representation power in comparison to MLP which can lead the density branch predicting fuzzy geometry to approximate the view-dependent effects.

\subsection{Surface parameterization}
Before the surface parameterization we simplify the mesh using vertex clustering algorithm.
We further remove the mesh faces that are invisible from the training views.
We also remove the regions of the faces where maximum volumetric weights across all training views are below a certain threshold, as this region belongs to the internal part of an object and does not contribute significantly to rendering.
We then segment the mesh into smaller patches using graph cuts.
Each segmented patch is parameterized separately and then all patches are packed into an atlas.
For mipNeRF 360 scenes, we parameterize the mesh in contracted space such that the triangles that are farther away use fewer texels per unit.

\subsection{Run-time performance}
We also implement rendering using the depth peeling algorithm~\cite{depthpeeling} to efficiently render in a browser while taking full advantage of hardware-accelerated rasterization, as shown in the \cref{tab:fpsmipnerf}.

\subsection{Effect of different mesh extraction techniques}
\begin{figure}
    \centering
    \includegraphics[width=\columnwidth]{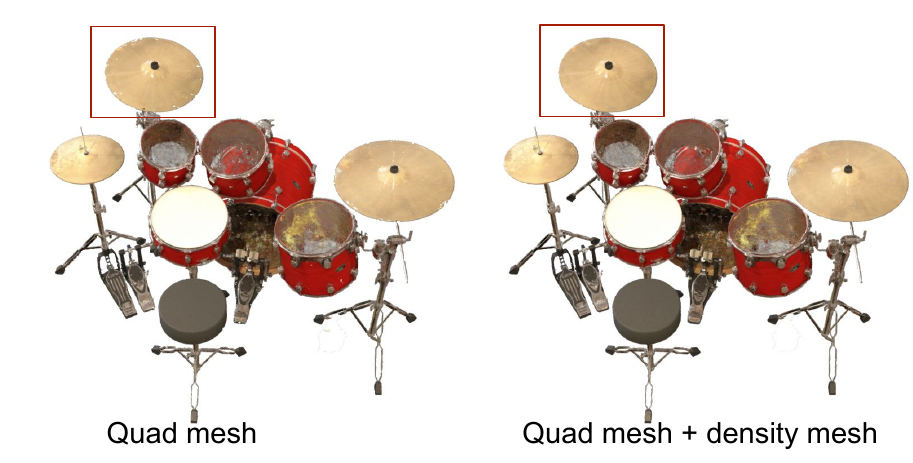}
    \caption{Mesh extracted from our quadrature field can result in holes. This can be alleviated by adding a coarse mesh extracted from the density field.}
    \label{fig:quad-mesh-holes}
\end{figure}

Mesh extracted from our quadrature field can result in holes. This can be alleviated by adding a coarse mesh extracted from the density field as shown in \cref{fig:quad-mesh-holes}. 
This comes at a little additional cost of more triangles per scene.

\begin{figure}
    \centering
    \includegraphics[width=0.8\linewidth]{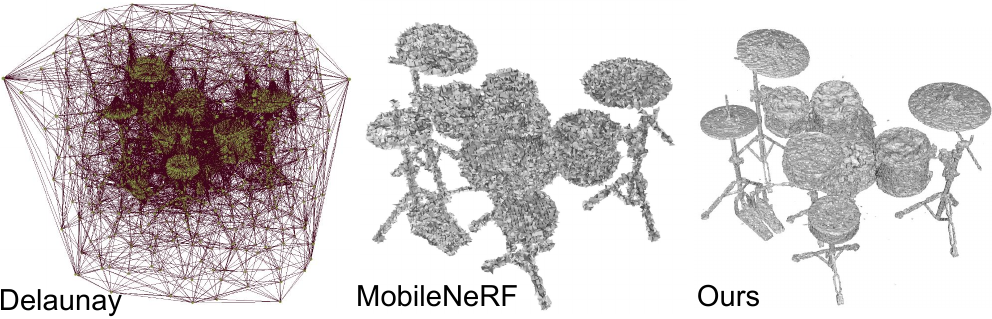}
    \caption{\textbf{Comparison of reconstructed geometry.} 
1)  Mesh reconstructed using Delaunay Triangulation on the surface field.
This approach produces more vertices of tetrahedra in regions where the surface is likely to occur.
The mesh reconstructed using Delaunay triangulation results in a bad representation of the underlying surface. 
2) Mesh reconstructed using MobileNeRF consists of polygonal soup.
3) Mesh reconstructed using our approach.}
\label{fig:mesh-comparison}
\end{figure}
\begin{table}
\centering
\resizebox{0.8\columnwidth}{!}{%
\begin{tabular}{lcc}
\toprule
                        & FPS (depth-peeling) & FPS (ray-tracing) \\
\hline
Desktop Nvidia-3090      &  15       &  140  \\
\bottomrule
\end{tabular}
}
\caption{Comparison of rendering speed between depth-peeling and ray-tracing implementations using mic scene of NeRF synthetic dataset at 2$\times$ super-resolution.
\label{tab:fpsmipnerf}
}
\end{table}

\subsection{Evaluation metrics}
We provide more evaluation using SSIM metric in \cref{tab:ssim-shelly}, \cref{tab:ssim-nerf-synthetic}, and \cref{tab:ssim-mipnerf360} and LPIPS  metric \cref{tab:lpips-shelly}, \cref{tab:lpips-nerf-synthetic} and \cref{tab:lpips-mipnerf360} respectively on Shelly, NeRF synthetic and mipNeRF 360 dataset.

\begin{table}[]
\centering
\caption{Evaluation of Shelly dataset using SSIM metric.}
\label{tab:ssim-shelly}
\begin{tabular}{lccccccc}
\toprule
SSIM               & Mean   & Fernvase & Pug    & Woolly & Horse  & Khady  & Kitten \\
\hline
Instant-NGP~\cite{mueller2022instant}                & 0.9675 & 0.9870   & 0.9634 & 0.9778 & 0.9947 & 0.8955 & 0.9870 \\
GS~\cite{kerbl3Dgaussians} & 0.9591 & 0.9871   & 0.9627 & 0.9470 & 0.9942 & 0.8793 & 0.9841 \\
MobileNeRF~\cite{chen2022mobilenerf}         & 0.9108 & 0.9440   & 0.8850 & 0.8910 & 0.9800 & 0.8230 & 0.9420 \\
Adaptive shells~\cite{adaptiveshells2023}    & 0.9542 & 0.9760   & 0.9470 & 0.9500 & 0.9920 & 0.8810 & 0.9790 \\
Ours    & 0.9545 & 0.9778   & 0.9385 & 0.9503 & 0.9901 & 0.8893 & 0.9813 \\
\bottomrule
\end{tabular}%
\end{table}
\begin{table}[]
\centering
\caption{Evaluation of Shelly dataset using LPIPS metric.}
\label{tab:lpips-shelly}
\begin{tabular}{lccccccc}
\toprule
LPIPS              & Mean   & Fernvase & Pug    & Woolly & Horse  & Khady  & Kitten \\
\hline
Instant-NGP~\cite{mueller2022instant}                & 0.0584 & 0.0256   & 0.0672 & 0.0545 & 0.0222 & 0.1475 & 0.0333 \\
GS~\cite{kerbl3Dgaussians} & 0.0670 & 0.0235   & 0.0737 & 0.0855 & 0.0244 & 0.1606 & 0.0345 \\
MobileNeRF~\cite{chen2022mobilenerf}         & 0.1288 & 0.0740   & 0.1670 & 0.1630 & 0.0570 & 0.2180 & 0.0940 \\
Adaptive shells~\cite{adaptiveshells2023}    & 0.0788 & 0.0460   & 0.0930 & 0.0890 & 0.0290 & 0.1600 & 0.0560 \\
%
Ours    & 0.0730 & 0.0369   & 0.0905 & 0.0950 & 0.0302 & 0.1400 & 0.0454 \\
\bottomrule
\end{tabular}
\end{table}

\begin{table}[]
\centering
\caption{Evaluation on NeRF synthetic dataset using SSIM metric.}
\label{tab:ssim-nerf-synthetic}
\resizebox{\textwidth}{!}{%
\begin{tabular}{lccccccccc}
\toprule
SSIM            & Mean   & lego   & chair  & ship   & mic    & drums  & materials & ficus  & hotdog \\
\hline
Instant-NGP~\cite{mueller2022instant}             & 0.9617 & 0.9803 & 0.9849 & 0.8943 & 0.9907 & 0.9328 & 0.9470    & 0.9819 & 0.9820 \\
Adaptive shells~\cite{adaptiveshells2023} & 0.9571 & 0.9730 & 0.9850 & 0.8770 & 0.9880 & 0.9370 & 0.9350    & 0.9810 & 0.9810 \\
MobileNeRF~\cite{chen2022mobilenerf}      & 0.9471 & 0.9750 & 0.9780 & 0.8670 & 0.9790 & 0.9270 & 0.9130    & 0.9650 & 0.9730 \\
Ours            & 0.9522 & 0.9687 & 0.9745 & 0.8787 & 0.9855 & 0.9274 & 0.9347    & 0.9764 & 0.9715 \\
\bottomrule
\end{tabular}
}
\end{table}


\begin{table}[]
\centering
\caption{Evaluation on NeRF synthetic dataset using LPIPS metric.}
\label{tab:lpips-nerf-synthetic}
\resizebox{\textwidth}{!}{%
\begin{tabular}{lccccccccc}
\toprule
LPIPS           & Mean & lego   & chair  & ship   & mic    & drums  & materials & ficus  & hotdog \\
\hline
Instant-NGP~\cite{mueller2022instant}             & 0.0518        & 0.0228 & 0.0215 & 0.1331 & 0.0152 & 0.0852 & 0.0736    & 0.0279 & 0.0349 \\
Adaptive shells~\cite{adaptiveshells2023} & 0.0563        & 0.0310 & 0.0230 & 0.1410 & 0.0150 & 0.0860 & 0.0860    & 0.0330 & 0.0350 \\
MobileNeRF~\cite{barron2022mipnerf360}      & 0.0618        & 0.0250 & 0.0250 & 0.1450 & 0.0320 & 0.0770 & 0.0920    & 0.0480 & 0.0500 \\
Ours            & 0.0691        & 0.0476 & 0.0405 & 0.1504 & 0.0262 & 0.0939 & 0.0871    & 0.0380 & 0.0690 \\
\bottomrule
\end{tabular}
}
\end{table}

\begin{table}[]
\centering
\caption{Evaluation of MipNeRF 360 dataset using SSIM metric.}
\label{tab:ssim-mipnerf360}
\resizebox{\textwidth}{!}{%
\begin{tabular}{lccccccccc}
\toprule
SSIM               & Mean indoors & KitchenLego & Room   & Bonsai & Kitchencounter & Mean outdoors & Garden & Bicycle & Stump  \\
\hline
Instant-NGP~\cite{mueller2022instant}                & 0.8663       & 0.8607      & 0.8839 & 0.9025 & 0.8182         & 0.6346        & 0.7270 & 0.5524  & 0.6245 \\
GS~\cite{kerbl3Dgaussians} & 0.9198       & 0.9220      & 0.9140 & 0.9380 & 0.9050         & 0.8047        & 0.8680 & 0.7710  & 0.7750 \\
MobileNeRF~\cite{chen2022mobilenerf}         &    -          &     -        &  -      &        &     -           & 0.5270        & 0.5990 & 0.4260  & 0.5560 \\
BakedSDF~\cite{yariv2023bakedsdf}           & 0.8365       & 0.8170      & 0.8700 & 0.8510 & 0.8080         & 0.6387        & 0.7510 & 0.5700  & 0.5950 \\
Adaptive Shells~\cite{adaptiveshells2023}    & 0.8723       & 0.8660      & 0.8950 & 0.9330 & 0.7950         & 0.6057        & 0.7570 & 0.5440  & 0.5160 \\
Ours            & 0.8598       & 0.8660      & 0.8763 & 0.8896 & 0.8075         & 0.6496        & 0.7520 & 0.5673  & 0.6296 \\
\bottomrule
\end{tabular}
}
\end{table}

\begin{table}[]
\centering
\caption{Evaluation on MipNeRF 360 dataset using LPIPS metric.}
\label{tab:lpips-mipnerf360}
\resizebox{\textwidth}{!}{%
\begin{tabular}{llllllllll}
\toprule
LPIPS              & Mean indoors & KitchenLego & Room   & Bonsai & Kitchencounter & Mean outdoors & Garden & Bicycle & Stump  \\
\hline
Instant-NGP~\cite{mueller2022instant}                & 0.2997       & 0.2328      & 0.3291 & 0.2801 & 0.3569         & 0.3849        & 0.2725 & 0.4602  & 0.4221 \\
GS~\cite{kerbl3Dgaussians} & 0.1895       & 0.1290      & 0.2200 & 0.2050 & 0.2040         & 0.1727        & 0.1030 & 0.2050  & 0.2100 \\
MobileNeRF~\cite{chen2022mobilenerf}         &      -        &      -       &    -    &    -    &         -       & 0.4337        & 0.3580 & 0.5130  & 0.4300 \\
BakedSDF~\cite{yariv2023bakedsdf}           & 0.2583       & 0.2370      & 0.2510 & 0.2590 & 0.2860         & 0.3173        & 0.2130 & 0.3680  & 0.3710 \\
Adaptive Shells~\cite{adaptiveshells2023}    & 0.2848       & 0.2190      & 0.3000 & 0.2250 & 0.3950         & 0.3890        & 0.2470 & 0.4380  & 0.4820 \\
Ours            & 0.3032       & 0.2315      & 0.3229 & 0.3023 & 0.3561         & 0.3513        & 0.2414 & 0.4220  & 0.3905 \\
\bottomrule
\end{tabular}
}
\end{table}

\subsection{Architecture design}
We implement a quadrature field using the hash grid with an MLP with two hidden layers, each of width 16.
The input coordinates are not only input to the hash grid but also concatenated with the output of the hash grid.
We use Exponential Linear Units~\cite{elu2015} in the MLP which allows double derivatives used to supervise our quadrature field.
The experiments on NeRF synthetic and Shelly dataset are done with the hash grid with a maximum resolution of 512, minimum resolution of 16 and codebook size of $2^{30}$.
For the real dataset, we use the hash grid with a maximum resolution of 4096, a minimum resolution of 16 and a codebook size of $2^{25}$.
We use a similar architecture design to implement our deformation network, except we use relu non-linearity in the MLP.
The choice of these parameters is dependent on available GPU memory and validation performance.

\begin{table}[]
\centering
\resizebox{0.7\columnwidth}{!}{%
\begin{tabular}{lccc}
\hline
         & \textbf{NeRF Synthetic} & \textbf{Shelly}      & \textbf{MipNeRF 360} \\
\hline
Disk (MB) & 328         & 1213 &    4331         \\
VRAM (MB)    & 803 & 2764 &    9066      \\
\hline
\end{tabular}%
}
\caption{Average memory requirement on different datasets.}\label{tab:storage}
\end{table}

\begin{figure}
\centering
\includegraphics[width=\columnwidth]{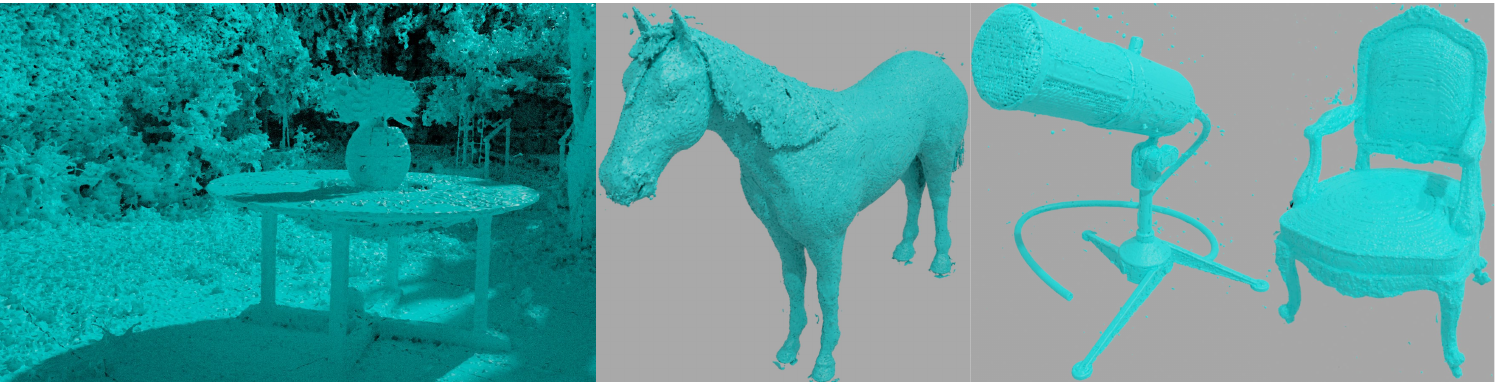}
\caption{\textbf{Geometry from our approach on different datasets.}}
\label{fig:geometry}
\end{figure}
\end{document}